\documentclass[prl,showpacs,showkeys,nofootinbib,twocolumn,floatfix]{revtex4}

\usepackage{graphicx}  %
\usepackage{amsmath}
\usepackage{bm}  %
\usepackage{ulem}

\newcommand{\bea}{\begin{eqnarray}}
\newcommand{\eea}{\end{eqnarray}}
\newcommand{\beq}{\begin{equation}}
\newcommand{\eeq}{\end{equation}}
\newcommand{\bqa}{\begin{eqnarray}}
\newcommand{\eqa}{\end{eqnarray}}

\def\mqo2{{\!\!\!}}

\begin{document}

\title{
Association of Atoms into Universal Dimers \\ 
using an Oscillating Magnetic Field}

\author{Christian Langmack}
\email{clangmack@gmail.com}
\affiliation{Department of Physics,
         The Ohio State University, Columbus, OH\ 43210, USA\\}

\author{D.~Hudson Smith}
\email{smith.7991@osu.edu}
\affiliation{Department of Physics,
         The Ohio State University, Columbus, OH\ 43210, USA\\}

\author{Eric Braaten}
\email{braaten@mps.ohio-state.edu}
\affiliation{Department of Physics,
         The Ohio State University, Columbus, OH\ 43210, USA\\}

\date{\today}

\begin{abstract}
In a system of ultracold atoms near a Feshbach resonance,
pairs of atoms can be associated into universal dimers 
by an oscillating magnetic field with frequency 
near that determined by the dimer binding energy.
We present a simple expression for the transition rate 
that takes into account many-body effects through 
a transition matrix element of the contact.
In a thermal gas, the  width of the peak in the transition rate
as a function of the frequency is determined by the temperature.
In a dilute Bose-Einstein condensate of atoms, 
the width is determined by the inelastic scattering rates of 
a dimer with zero-energy atoms.
Near an atom-dimer resonance, there is a dramatic increase 
in the width from inelastic atom-dimer scattering
and from atom-atom-dimer recombination.
The recombination contribution provides a signature for 
universal tetramers that are Efimov states 
consisting of two atoms and a dimer.
\end{abstract}

\smallskip
\maketitle

{\bf Introduction}.
The field of ultracold atoms has extended the frontiers of few-body 
and many-body physics by providing pristine systems in which the interactions 
between the constituents are simple and have strengths that can be controlled 
experimentally.  The frontiers of few-body physics include the study
of universal molecules, which have properties 
determined by the large scattering length $a$ of the atoms,
and universal reaction rates, whose dependence on $a$ and on 
kinematic variables is consistent with asymptotic scale invariance
or discrete scale invariance \cite{Braaten:2004rn}.
The need for accurate calculations of universal properties has pushed 
the computational frontiers to the 4-body problem and beyond.
In many-body physics, the frontiers include the study of superfluidity
and other novel phases of matter \cite{IKS:0801}.  
Accurate measurements of the properties of 
systems of ultracold atoms
present a challenge to many-body calculational methods
because of the strong correlations produced 
by a large scattering length.
Particularly challenging is the unitary limit in which $a$
is infinitely large and the interactions between the constituents 
are the strongest allowed by quantum mechanics.

Few-body physics provides powerful constraints on many-body physics 
through universal relations pioneered by Tan
\cite{Tan0505,Tan0508,Tan0803}.
Many of these relations involve the {\it contact}, 
an extensive property of the system that is 
conjugate to $1/a$ and provides a measure of 
the probability for pairs of particles to be very close together.
The contact controls the thermodynamics of a many-body system
and also determines the high-momentum and high-frequency tails 
of correlation functions \cite{Braaten:2010if}.  

One way to produce universal molecules in a system of ultracold atoms is 
magneto-association -- the modulation of the magnetic field 
near a Feshbach resonance with frequency 
near that determined by the binding energy of the molecule.
This method was first used by Thompson, Hodby, and Wieman
to produce shallow dimers composed of $^{85}$Rb atoms  \cite{Wieman0505}.
It has since been used to produce dimers with various other atoms 
as constituents and to measure their binding energies
\cite{Wieman0607,Jin0703,Jin0712,Inguscio0808,Grimm0810,Khaykovich1009,Hulet1302}.
Magneto-association has also been used to associate $^7$Li atoms 
into Efimov trimers \cite{Khaykovich1201}.

In this paper, we derive the magneto-association rate 
for universal molecules in a many-body system of ultracold atoms.
Many-body effects are taken into account
through a transition matrix element of the contact operator.
We deduce simple expressions for the transition rate for
producing universal dimers in a
thermal gas of bosons or fermions and in a Bose-Einstein condensate (BEC)
as a function of frequency.
The dramatic increase in the width of the peak in the transition rate 
near an atom-dimer resonance provides a signature 
for new universal tetramers that are
Efimov states consisting of two atoms and a dimer.

{\bf Transition rate}.
We consider a system of ultracold atoms in a magnetic field 
that has a constant value $\bar B$ for $t<0$
and oscillates with a small amplitude $b$ for $t>0$:
$B(t) = \bar B + b \sin(\omega t)$.
Near a Feshbach resonance, 
the scattering length is a function of the magnetic field: 
$a(B) = a_{\rm bg}[1 - \Delta/(B - B_0)]$,
where $a_{\rm bg}$ is the background scattering length,
and $B_0$ and $B_0 + \Delta$ are the positions of the pole 
and the zero of the scattering length, respectively. 
The inverse scattering length can be expanded in powers of $b$:
\begin{equation}
\frac{1}{a(t)} = 
\frac{1}{\bar a} 
- \frac{\Delta b}{a_{\rm bg}(\bar B - B_0 - \Delta)^2} \sin(\omega t) 
+ \ldots ,
\label{a-inverse}
\end{equation}
where $\bar a = a(\bar B)$.
The deviation of $1/a(t)$ from $1/\bar a$ can be treated as
a periodic time-dependent perturbation.  By Tan's adiabatic relation, 
a small change in $1/a$ produces a change in the energy 
that is proportional to the {\it contact} $C$ \cite{Tan0508}.
Thus the perturbing Hamiltonian is proportional to the contact.
In the case of identical bosons with mass $m$, it can be expressed as
\begin{equation}
H_{\rm pert}(t) = 
- \frac{\hbar^2}{8 \pi m}
\left(\frac{1}{a(t)}  - \frac{1}{\bar a} \right) C .
\label{H-C}
\end{equation}
(In the case of fermions with two spin states,
the prefactor should be multiplied by 2.)
The leading term of order $b$ in $H_{\rm pert}(t)$
drives transitions to states with energies that are 
higher or lower by $\hbar \omega$.
Higher order terms drive transitions to states 
whose energies differ by larger integer multiples of $\hbar \omega$.
If $\bar a$ is large and positive 
and if $\hbar \omega$ is near the binding energy 
$\hbar^2/m \bar a^2$ of the shallow dimer,
the first-order perturbation can associate pairs of atoms into dimers.

The transition rate $\Gamma(\omega)$ of the initial state $| i \rangle$
into final states $| f \rangle$
at leading order in $b$ is given by Fermi's Golden Rule:
\begin{eqnarray}
\Gamma(\omega)  = 
\frac{\hbar^3 \Delta^2 b^2}
    {256 \pi^2 m^2 a_{\rm bg}^2 (\bar B - B_0 - \Delta)^4}
\sum_f \big| \langle f | C | i \rangle \big|^2
\nonumber
\\
\times \sum_\pm 
\frac{\hbar \Gamma_f}{|E_i \pm \hbar \omega - E_f + i \hbar \Gamma_f/2|^2} ,
\label{Gamma-omega}
\end{eqnarray}
where $E_i$ and $E_f$ are the energies of the initial state
and the final states, respectively.
(In the case of fermions with two spin states,
the prefactor should be multiplied by 4.)
The Lorentzian factor allows for the possibility that the final state 
involves the excitation of a resonance
with lifetime $1/\Gamma_f$.
In the limit $\Gamma_f \to 0$, this factor reduces to 
$2 \pi \delta(E_i \pm \hbar \omega - E_f)$.

The association of molecules in a time-dependent magnetic field 
has been considered previously by Hanna, K\"ohler, and Burnett
\cite{HKB0609}.  They calculated the probability for producing a dimer 
as a function of time by solving the time-dependent Schroedinger equation 
for two atoms in a two-channel model with a closed channel.
A major disadvantage of this approach is its inability to account 
for many-body effects, 
which are taken into account in Eq.~(\ref{H-C})
through the transition matrix element of $C$.

It is convenient to express the contact operator in Eq.~(\ref{H-C})
as the integral of the {\it contact density} operator:
$C = \int d^3r \, {\cal C}(\bm{r})$.
The field theoretic definition of the contact \cite{Braaten:2008uh}
reveals that the contact density operator can be expressed 
as ${\cal C} = \phi^\dagger \phi$,
where the {\it contact field} $\phi(\bm{r})$ is a local operator 
that annihilates two atoms at a point.
The transition matrix element can be expressed as
\begin{equation}
\langle f | C | i \rangle = \int \!d^3r \, 
\langle f | \phi^\dagger(\bm{r}) \, \phi(\bm{r}) | i \rangle .
\label{<C>-phi} 
\end{equation}
A complete set of states $\sum_n | n \rangle \langle n | = 1$
can be inserted between $\phi^\dagger$ and $\phi$.
If only one term in the sum is nonzero, 
the matrix element factors into a matrix element of $\phi$
that involves the initial state 
and a matrix element of $\phi^\dagger$
that involves the final state. 

For a many-body system whose number density $n(\bm{R})$
varies slowly with the position $\bm{R}$,
the transition rate can be simplified by using the
local density approximation.
The matrix element of ${\cal C}$ can be expressed 
in terms of the matrix element for the homogeneous system 
whose initial state $| i \rangle$ has constant
number density equal to $n(\bm{R})$.
By exploiting the translational invariance 
of the homogeneous system, the modulus squared of the matrix element 
summed over final states can be reduced to
\begin{eqnarray}
\sum_f \big| \langle f | C | i \rangle \big|^2 &=&
\sum_f (2 \pi \hbar)^3 \delta^3(P_i - P_f)
\nonumber
\\
&& \times \int \! d^3R  \,
\big| \langle f | \phi^\dagger(\bm{R}) \, \phi(\bm{R}) | i \rangle \big|^2,
\label{<C>-LDA}
\end{eqnarray}
where $P_i$ and $P_f$ are the total momenta of the initial 
and final states of the homogeneous system, respectively.

In a system of ultracold trapped bosonic atoms, 
a low-momentum dimer produced by magneto-association 
will eventually suffer an inelastic collision,
producing energetic particles that escape from the 
trapping potential.  An inelastic collision with a single atom 
results in the loss of 3 atoms.
An inelastic collision with two atoms 
results in the loss of 4 atoms.
The transition rate can be determined
from measurements of the loss of trapped atoms.
For fermionic atoms, a different method would be required
to measure the transition rate.

{\bf Thermal gas}.
We first consider a dilute thermal gas of atoms, 
whose momentum distribution can be approximated 
by a Boltzmann distribution with temperature $T$
and local number density $n$.
If the gas consists of a large number $N$ of bosonic atoms,
any of the $N^2/2$ pairs of atoms can make the transition to the dimer.
The transition matrix element 
reduces to the matrix element of $\phi^\dagger\phi$ between the 
dimer state $\langle {\rm D} |$ and the state $| {\rm A A} \rangle$ 
for the pair of atoms that makes the transition.
Upon inserting the projector $| 0 \rangle \langle 0 |$ 
onto the vacuum state between $\phi^\dagger$ and $\phi$,
the square of the matrix element can be expressed as the product
of the contact for a dimer, which is equal to $16 \pi/\bar a$,
the contact for a pair of atoms with relative wavenumber $k$, 
which can be deduced from the contact 
for a pair of atoms in thermal equilibrium derived in 
Ref.~\cite{Braaten:2013eya}, and factors of the volume $V$ 
associated with the normalization of plane-wave states:
\begin{equation}
\big| \langle {\rm D} | \phi^\dagger\phi | {\rm A A} \rangle \big|^2 =
\frac{1024 \pi^3 \bar a}{(1 + \bar a^2 k^2) V^2}.
\label{<C>-thermal}
\end{equation}
The factor of $N^2/V^2$ can be replaced by $n^2(\bm{R})$.
The Lorentzian function in Eq.~(\ref{Gamma-omega})
reduces to a delta function.
The transition rate is zero if $\hbar \omega < \hbar^2/m \bar a^2$.
For larger $\omega$, the transition rate is 
\begin{eqnarray}
\Gamma(\omega)  &=& 
\frac{\sqrt{2} \hbar \Delta^2 b^2 \bar a }
{m a_{\rm bg}^2 (\bar B - B_0 - \Delta)^4} 
\left( \int \! d^3R \, n^2(\bm{R}) \right)
\nonumber
\\
&& \times 
\frac{\lambda_T^3  k(\omega)}{1 + \bar a^2  k^2(\omega)} 
\exp(- \lambda_T^2 k^2(\omega)/2 \pi) ,
\label{Gamma-thermal}
\end{eqnarray}
where $k(\omega)= (m \omega/\hbar - 1/\bar a^2)^{1/2}$
and $\lambda_T = (2 \pi \hbar^2/mkT)^{1/2}$.
(The transition rate for fermions is obtained by replacing
$n^2$ by $2 n_1 n_2$, where $n_i$ is the number density for spin state $i$.)
The first subharmonic transition rate due to the order-$b^2$
perturbation in Eq.~\eqref{H-C}
can be obtained from Eq.~\eqref{Gamma-thermal}
by multiplying by $[b/2(\bar B -B_0 - \Delta)]^2$
and making the substitution $\omega \to 2 \omega$.

\begin{figure}[t]
\centerline{ \includegraphics*[width=8.6cm,clip=true]{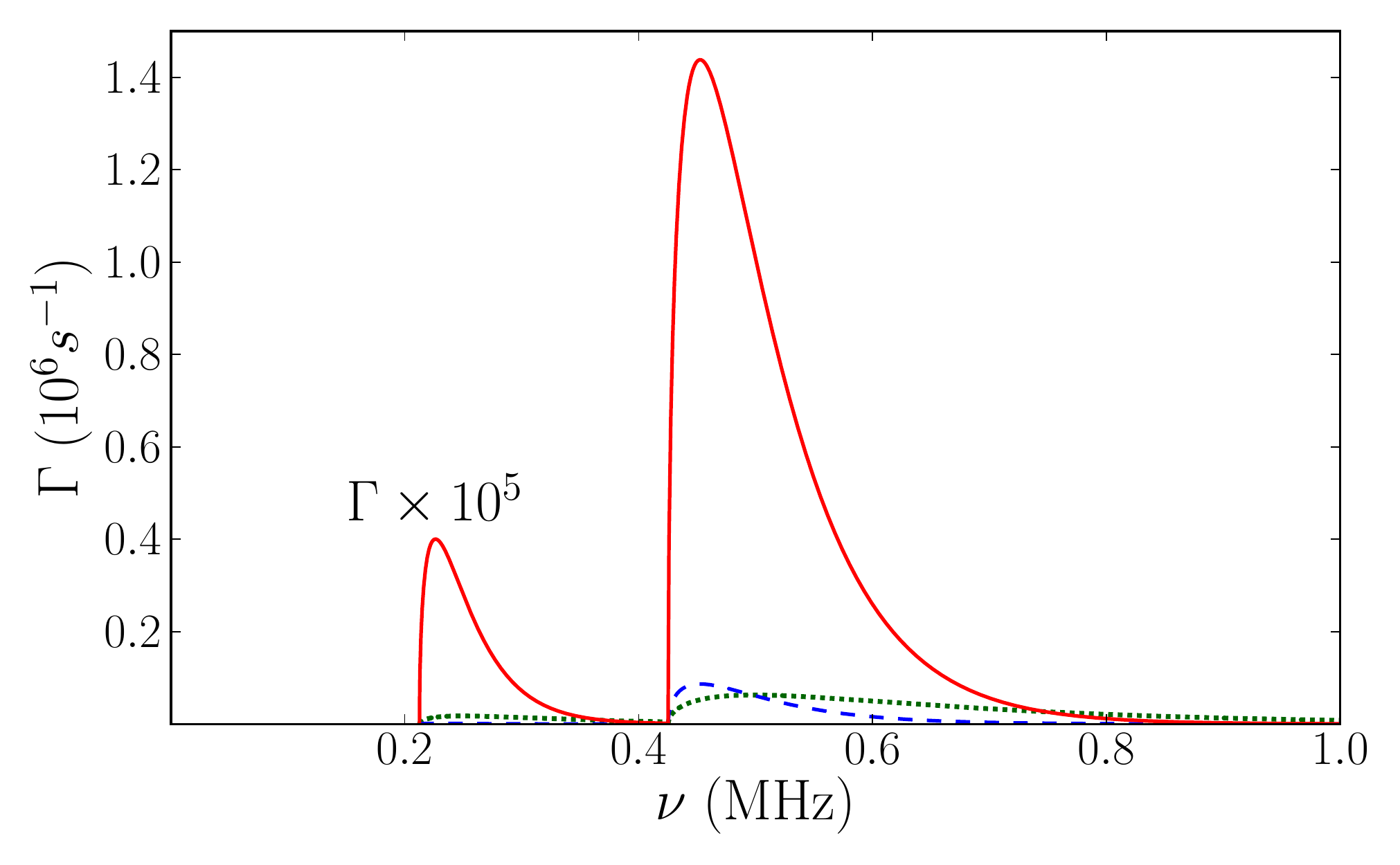} }
\vspace*{0.0cm}
\caption{Transition rate $\Gamma$ for producing dimers
in a thermal gas of $^7$Li atoms at $\bar B = 734.5$~G
as a function of the frequency $\nu$.
The curves are the transition rate in Eq.~\eqref{Gamma-thermal} 
and the first subharmonic transition rate 
for three combinations of the 
modulation amplitude $b$ and the temperature $T$:
$(b,T) = (0.57~G,10~\mu K)$ (dotted, green),
$(0.14~G,3~\mu K)$ (dashed, blue),
$(0.57~G,3~\mu K)$ (solid, red).
}
\label{fig:Gammathermal}
\end{figure}

The loss of atoms from magneto-association into dimers 
has been studied by Dyke {\it et al.}\ 
using a thermal gas of $^7$Li atoms
at a magnetic field $\bar B = 734.5$~G, 
where the scattering length is $\bar a = 1100~a_0$ \cite{Hulet1302}.
They reported the fraction of atoms remaining 
after an unspecified time as a function of the frequency
for three combinations of the modulation amplitude and the temperature.
In Fig.~\ref{fig:Gammathermal},
the predicted transition rates $\Gamma(\omega)$ 
for these three conditions, including the first subharmonic,
are shown as functions of the frequency $\nu =\omega/2 \pi$.

{\bf Bose-Einstein condensate}.
We next consider a dilute BEC of atoms
at zero temperature with local number density $n$.
The contact field $\phi$ can be expressed as the sum of 
its expectation value $\bar \phi$ and a quantum fluctuation field 
$\tilde \phi$:  $\phi(\bm{r}) = \bar \phi + \tilde \phi(\bm{r})$.  
The field $\tilde \phi^\dagger$ has a nonzero amplitude 
to create a dimer in the BEC.
The contribution to the matrix element of $\phi^\dagger \phi$ 
between the state  $\langle i+{\rm D} |$ 
in which a dimer has been excited and the BEC $| i \rangle$
comes from the $\tilde \phi^\dagger \bar \phi$ term.
The product of $\bar \phi$ and its complex conjugate 
is the contact density of the BEC, which in the dilute limit is
$\bar \phi^* \bar \phi = 16 \pi^2 \bar a^2 n^2$. 
The square of $\langle i+{\rm D} | \tilde \phi^\dagger | i \rangle$
is the contact of the dimer, which in the dilute limit
is simply $16 \pi/\bar a$.
The most dramatic dependence on the frequency comes from the 
Lorentzian in  Eq.~(\ref{Gamma-omega}).
The dimer in the BEC behaves like a resonance whose complex energy 
$E_{\rm D} - i \hbar \Gamma_{\rm D}/2$ is the
sum of the binding energy $-\hbar^2/m \bar a^2$
and the mean field energy from forward scattering of the dimer 
from 0-momentum atoms in the condensate.
The transition rate to dimers in the BEC is
\begin{eqnarray}
\Gamma(\omega)  &=& 
\frac{\pi \hbar^3 \Delta^2 b^2 \bar a }
       {m^2 a_{\rm bg}^2 (\bar B - B_0 - \Delta)^4}
 \nonumber
 \\
&& \times 
\int \! d^3R \; n^2(\bm{R})
\frac{\hbar \Gamma_{\rm D}}
    {(E_{\rm D} + \hbar \omega)^2 + \hbar^2 \Gamma_{\rm D}^2/4} .
\label{Gamma-BEC}
\end{eqnarray}
The spatial integral is a density-weighted average of a Lorentzian 
with a density-dependent width $\Gamma_{\rm D}$.

The complex energy of a dimer in the BEC is given by
\begin{subequations}
\begin{eqnarray}
E_{\rm D}  &=& 
- \frac{\hbar^2}{m a^2} 
+ \frac{3 \pi \hbar^2}{m} ({\rm Re} \, a_{\rm AD})\, n + \ldots,
\label{E-D}
\\
\Gamma_{\rm D}  &=& \beta_{\rm AD} n 
+ \mbox{$\frac12$} \alpha_{\rm AAD} n^2 + \ldots.
\label{Gamma-D}
\end{eqnarray}
\label{E,Gamma-D}%
\end{subequations}
The leading mean-field correction to $E_{\rm D}$
comes from the atom-dimer (AD) scattering length $a_{\rm AD}$.
By the optical theorem, $\Gamma_{\rm D}$
is determined by the inelastic scattering rate of the dimer.
If there are deep dimers, the leading contribution to $\Gamma_{\rm D}$
comes from AD scattering into an atom and a deep dimer,
whose rate coefficient is
$\beta_{\rm AD} = 6 \pi \hbar(-{\rm Im} \, a_{\rm AD})/m$.
The universal results for $a_{\rm AD}$ and $\beta_{\rm AD}$
are given in Ref.~\cite{Braaten:2004rn} as functions of $a_*$ and $\eta_*$,
where $a_* \exp(-i \eta_*/s_0)$ is the complex scattering length
where $a_{\rm AD}$ diverges.
In the absence of deep dimers, the leading contributions to $\Gamma_{\rm D}$
come from atom-atom-dimer (AAD) recombination into two shallow dimers 
or into an Efimov trimer and an atom.
We define $\alpha_{\rm AAD}$
so that the event rate per volume in a dilute thermal gas with 
atom and dimer number densities $n$ and $n_D$
is $\alpha_{\rm AAD} n^2 n_D$.

{\bf Atom-atom-dimer recombination}.
The universal result for the AAD recombination rate 
at threshold has been calculated by Deltuva \cite{Deltuva:2012xf}.
The rate constant $\alpha_{\rm AAD}$ 
can be expressed as $\hbar a^4/m$
multiplied by a log-periodic function of $a$ 
with discrete scaling factor $e ^{\pi/s_0} \approx 22.7$,
where $s_0  \approx 1.00624$.
The log-periodic function has a minimum 
value of $118$ at $a = 5.5~a_*$.
It increases dramatically 
as $a$ approaches $a_*$ as a consequence of 
the Efimov effect in the AAD system.
The associated Efimov states are universal tetramers whose existence 
was first pointed out by Braaten and Hammer \cite{Braaten:2004rn}.
Their binding energies and widths in the zero-range limit
have been calculated by Deltuva \cite{Deltuva:2012ms}.
The constituents of the AAD system have a mass ratio of 2
and the most resonant interaction is between an atom and a dimer.
The associated discrete scaling factor for $a_{\rm AD}$ is 
$e ^{\pi/s_2} \approx 2 \times 10^5$, where $s_2 \approx 0.257206$.
The universal 3-body recombination rates for systems 
with two identical particles that have resonant interactions 
with a third particle have been calculated by Helfrich,
Hammer, and Petrov for arbitrary mass ratio \cite{Helfrich:2010yr}.
Near an AD resonance $a_*$ where $a_{\rm AD} = \pm \infty$, 
the limiting behavior of $\alpha_{\rm AAD}$ is 
$(k_{\rm AT} + k_{\rm DD}) \hbar a_{\rm AD}^4/m$,
where $k_{\rm AT} + k_{\rm DD}$ is a log-periodic function 
of $a_{\rm AD}$ with discrete scaling factor $e ^{\pi/s_2}$.
For $a$ approaching $a_*$ from above,
the separate contributions from final states consisting 
of shallow trimer and atom or two dimers are
\begin{subequations}
\begin{eqnarray}
k_{\rm AT}  =
\frac{2 k_2 \big( \sin^2[s_2 \log(a_{\rm AD}/a_{2+})] + \sinh^2 \eta_{2*} \big)}
{\sinh^2(\pi s_2 + \eta_{2*}) + \cos^2[s_2 \ln(a_{\rm AD}/a_{2+})]} ,
\label{kAT}
\\
k_{\rm DD}  = 
\frac{(k_2/\tanh(\pi s_2)) \sinh(2\eta_{2*})}
{\sinh^2(\pi s_2 + \eta_{2*}) + \cos^2[s_2 \ln(a_{\rm AD}/a_{2+})]} .
\label{kDD}
\end{eqnarray}
\label{kATkDD}%
\end{subequations}
The atom-trimer contribution has interference minima
when $a_{\rm AD}$ is equal to $(e^{\pi/s_2})^n a_{2+}$,
where $n$ is an integer.
For $a$ approaching $a_*$ from below,
the combined contribution from final states consisting 
of two dimers or deeper trimer and atom are
\begin{eqnarray}
k_{\rm DD} + k_{\rm AT} =
\frac{(k_2/\tanh(\pi s_2)) \sinh(2\eta_{2*})}
{\sin^2[s_2 \ln(a_{\rm AD}/a_{2-})] + \sinh^2 \eta_{2*}} .
\label{kATDD}
\end{eqnarray}
There are resonance peaks when $a_{\rm AD}$ is equal to 
$(e^{\pi/s_2})^n a_{2-}$, where $n$ is an integer,
from Efimov states passing through the AAD threshold.
The coefficient $k_2 = 36.3367$
and the ratio $|a_{2-}|/a_{2+} = e^{\pi/2 s_2} \approx 449.053$ 
are universal constants.
Efimov states disappear through the atom-trimer threshold
when $a_{\rm AD}$ is equal to 
$(e^{\pi/s_2})^n a_{2*}$, where $n$ is an integer.
The universal ratio $a_{2*}/|a_{2-}| \approx 0.90$ 
can be determined by interpolating between numerical results 
given in Ref.~\cite{Helfrich:2010yr}.
The value of $a_{2*}$ was determined by Deltuva in 
Ref.~\cite{Deltuva:2012ms}: $a_{2*} \approx 1.608~a_*$.
The value of $\eta_{2*}$ can be determined by fitting
Deltuva's results in Ref.~\cite{Deltuva:2012xf}: $\eta_{2*} \approx 0.01$.
If there are deep dimers, they provide additional recombination channels.
Their effects can be taken into account 
by making the substitution 
$a_{*} \to a_{*} e^{-i\eta_*/s_0}$
in the amplitudes that give the rate constants 
in Eqs.~\eqref{kATkDD} and \eqref{kATDD}.

{\bf Dimer width in the BEC}.
In a dilute BEC of trapped atoms, the transition rate 
for producing dimers given by Eq.~(\ref{Gamma-BEC})
has a peak for $\omega$ near $\hbar/ma^2$.  
For a generic scattering length, 
the effect of the AD scattering term in Eq.~(\ref{E-D}) 
is to shift the peak by a fractional amount of order $na^3$,
which is small if the BEC is dilute. 
Near an AD resonance $a_*$, the fractional shift
increases to order $n a^2 |a_{\rm AD}|$.
However the fractional shift from AAD scattering
is of order $n^2 a^2 |a_{\rm AD}|^4$, which can be larger 
if $n |a_{\rm AD}|^3$ is much larger than 1.

The width $\Gamma_{\rm D}$ of the peak in the transition rate 
is given by Eq.~(\ref{Gamma-D}).
For a generic scattering length, the contributions to $\Gamma_{\rm D}$
from inelastic AD scattering and from AAD 
recombination are suppressed relative to $\hbar^2/ma^2$ 
by factors of order $\eta_* n a^3$ and $n^2 a^6$, respectively.
Near an AD resonance $a_*$, these factors 
increase to order $\eta_* n a |a_{\rm AD}|^2$ 
and $n^2 a^2 |a_{\rm AD}|^4$, respectively.
When $n a |a_{\rm AD}|^2$ is much larger than $\eta_*$,
the AAD contribution to $\Gamma_{\rm D}$ can be 
larger than the AD contribution.

\begin{figure}[t]
\centerline{ \includegraphics*[width=8.6cm,clip=true]{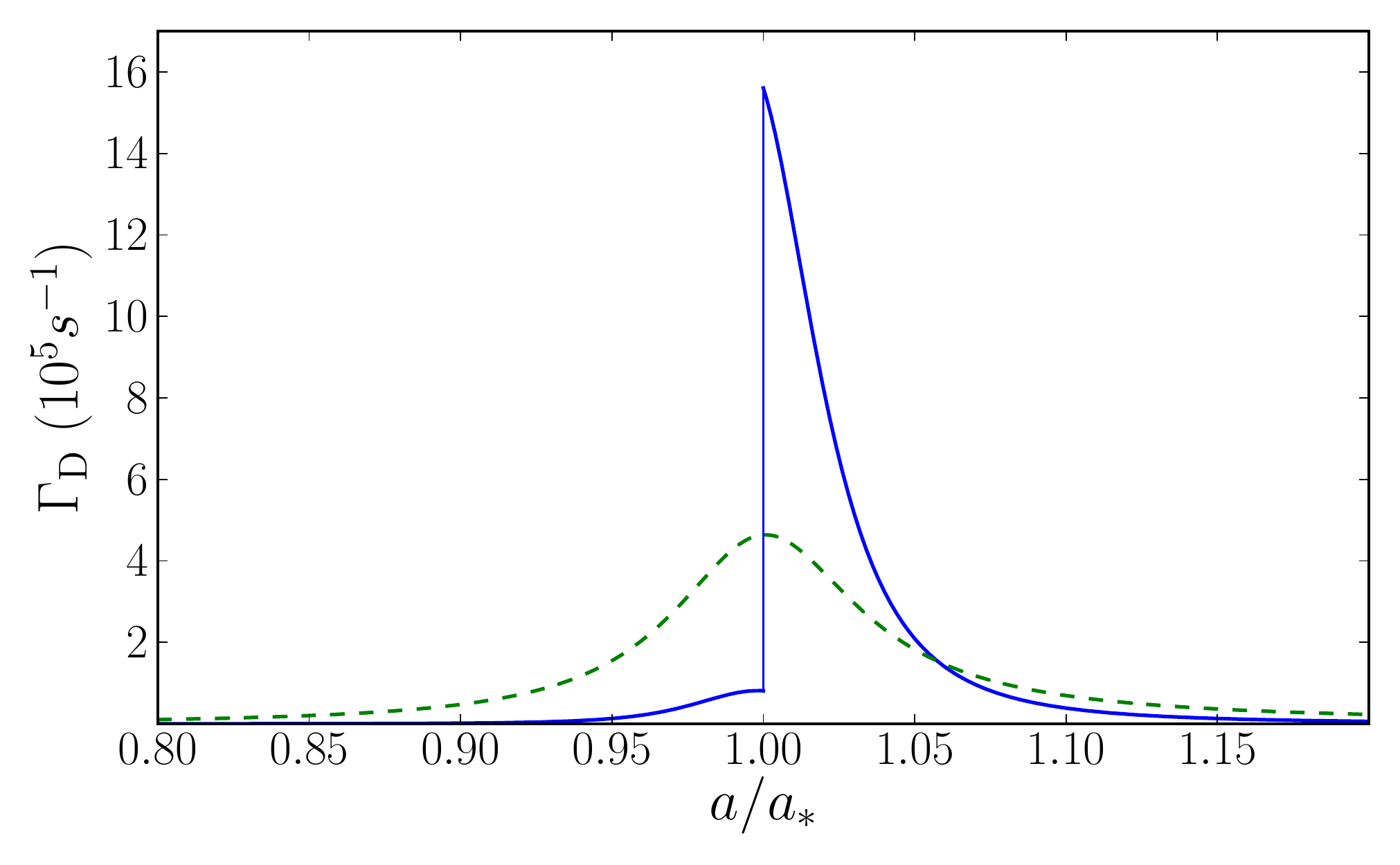} }
\vspace*{0.0cm}
\caption{Interaction width $\Gamma_{\rm D}$ of a dimer 
in a BEC of $^7$Li atoms with number density $n = 2.85/\mu m^3$
as a function of the scattering length $a$.
The curves are the contributions to $\Gamma_{\rm D}$
from inelastic atom-dimer collisions (dashed, green) and 
from atom-atom-dimer recombination (solid, blue).
}
\label{fig:Gamma-dimer}
\end{figure}

In Ref.~\cite{Hulet1302}, Dyke {\it et al.}\
also studied the loss of atoms from their magneto-association 
into dimers in a BEC of about $4 \times 10^5$ $^7$Li atoms
at a magnetic field $\bar B = 734.5$~G
with modulation amplitude $b = 0.14$~G.
The local number density can be approximated
by a Thomas-Fermi density profile with 
central number density $n = 2.85/\mu m^3$. 
In Fig.~\ref{fig:Gammathermal},
the dimer width $\Gamma_{\rm D}$ for
$a_* = 317~a_0$ and $\eta_* = 0.038$
is shown as a function of $a$.
The AD contribution is a Lorentzian centered at $a_*$.
The AAD contribution is discontinuous at $a_*$,
because there is an additional contribution for $a > a_*$
from recombination into an atom and the Efimov trimer
that disappears through the atom-dimer threshold at $a_*$.
The AAD contribution is smaller than the AD contribution 
for $a < a_*$, but it is larger 
in the range $a_* < a < 1.05~a_*$.
If $n$ is changed by a factor of $x$, 
the AD and AAD contributions
change by factors of $x$ and $x^2$, respectively.

{\bf Summary.}
We have derived a simple expression 
for the magneto-association rate of universal molecules 
that takes into account many-body effects 
through the transition matrix element of the contact.
We have applied it to the magneto-association of 
atoms into dimers in a thermal gas and in a BEC.
The width of the dimer peak in a BEC 
is dramatically enhanced near an atom-dimer resonance.
The contribution to the width from atom-atom-dimer 
recombination provides a signature for universal tetramers 
that are Efimov states consisting of two atoms and a dimer.
There are many other applications of 
the transition rate in an oscillating magnetic field,
including the magneto-association of atoms into Efimov trimers 
and the magneto-dissociation of paired fermions in a superfluid.
Thus the transition rate in an oscillating magnetic field 
provides a new window into the constraints on many-body physics
provided by few-body physics.

\newpage

\begin{acknowledgments}
This research was supported in part by the
National Science Foundation under grant PHY-1310862
and by the Simons Foundation.
This project was initiated during a workshop 
at the Institute for Nuclear Theory.
The possibility of observing the effects of 
atom-atom-dimer recombination though the width of the peak 
in the magneto-association rate was suggested by Randy Hulet.

\end{acknowledgments}

\end{document}